\documentclass[twocolumn,preprint2]{aastex631}
\usepackage{epsfig}
\usepackage{graphicx}
\usepackage{amsmath}
\usepackage{amsfonts}
\usepackage{soul}

\submitjournal{ApJ}

\begin{document}

\title{Reconstruction of photospheric velocity fields from highly corrupted data}

\correspondingauthor{Erico L. Rempel}
\email{rempel@ita.br}

\author{Erico L. Rempel}
\affiliation{Aeronautics Institute of Technology (ITA), 
S\~ao Jos\'e dos Campos, SP 12228–900, Brazil}
\affiliation{National Institute for Space Research (INPE), 
P.O. Box 515, S\~ao Jos\'e dos Campos, SP 12227-010, Brazil}
\author{Roman Chertovskih}
\affiliation{Research Center for Systems and Technologies (SYSTEC), 
Electrical and Computer Engineering Department, 
Faculty of Engineering, University of Porto, Rua Dr. Roberto Frias, s/n 4200-465 Porto, Portugal}
\author{Kamilla R. Davletshina}
\affiliation{Yandex, Lva Tolstogo street, 16, 119021 Moscow, Russian Federation}
\author{Suzana S. A. Silva}
\affiliation{Plasma Dynamics Group, Department of Automatic Control and Systems Engineering, University of Sheffield, Sheffield, UK}
\author{Brian T. Welsch}
\affiliation{University of Wisconsin, Green Bay, WI 54311, USA}
\author{Abraham C.-L. Chian}
\affiliation{National Institute for Space Research (INPE), 
P.O. Box 515, S\~ao Jos\'e dos Campos, SP 12227-010, Brazil}
\affiliation{University of Adelaide, School of Mathematical Sciences, Adelaide, SA 5005, Australia}

\date{\today}

\begin{abstract}
The analysis of the photospheric velocity field is essential for understanding plasma turbulence in the solar surface, which may be responsible for driving processes such as magnetic reconnection, flares, wave propagation, particle acceleration, and coronal heating. Currently, the only available methods to estimate velocities at the solar photosphere transverse to an observer's line of sight infer flows from differences in image structure in successive observations. Due to data noise, algorithms such as local correlation tracking (LCT) may lead to a vector field with wide gaps where no velocity vectors are provided.
In this letter, a novel method for image inpainting of highly corrupted data is proposed and applied to the restoration of horizontal velocity fields in the solar photosphere. The restored velocity field preserves all the vector field components present in the original field. The method shows robustness when applied to both simulated and observational data. 
\end{abstract}



\section{Introduction}


Nonlinear phenomena taking place in the solar photosphere  can strongly impact the plasma in the solar chromosphere and corona. Consider, for example,  the problem of coronal heating, wherein the temperature of the solar atmosphere is observed to increase drastically, from a few thousand degrees Kelvin to over 1 million degrees Kelvin, across a thin ($
\sim$ 100 km) transition region \citep{vernazza81}. It has been attributed to the excitation and propagation of Alfv\'en waves that transport energy from the photosphere to the upper solar atmosphere, and these magnetohydrodynamic (MHD) waves can be excited by swirling motions in the photospheric and chromospheric plasmas \citep{liu19,wedemeyer12}. Alternatively, coronal heating may be due to the occurrence of  nanoflares in the solar atmosphere \citep{parker88,testa14,bahauddin21}. Plasma turbulence in the photosphere and corona can also be responsible for magnetic reconnection events that may lead to strong solar flares and coronal mass ejections, with significant effects on space weather through the solar wind \citep{moore18,kusano20}. Therefore, a proper understanding, and possibly the capability to predict such phenomena through data-driven numerical simulations, depend on knowledge of the plasma motions in the photosphere. With that goal, different methods have been proposed to reconstruct the photospheric velocity field from available image sequences. Usually, a time series of observations of line-of-sight magnetogram, continuum intensity or dopplergram is employed to detect the motion of magnetic structures through some local correlation tracking (LCT) method \citep{november88,berger98,welsch04}, of which one of the most widely used is the Fourier (FLCT) method \citep{welsch04,fisher08,yeates12,chian14,liu19,birch19}. Such methods search for strong correlations between intensity features in image sequences to obtain velocity vectors. 

Despite their success in reconstructing photospheric velocity fields from available magnetograms, the LCT methods frequently suffer from data noise \citep{welsch12} or insufficient image structure \citep{schuck06}. In general, noisy fluctuations in regions of weak magnetic field lead to spurious correlations, thus, reconstructed velocity vectors are typically discarded where the line-of-sight magnetic field ($B_{{LOS}}$) is below a certain threshold. This may result in wide gaps in the reconstructed velocity fields that prevent their use as inputs in numerical simulations, for example. 
This problem is not limited to the LCT method; essentially all optical flow estimation methods assume that temporal variations in intensity from one image to the next arise from velocities transporting matter.   If, however, part of intensity fluctuations are spurious --- due, for instance, to measurement noise --- then the resulting flow estimates will also contain spurious components. While such unphysical flows might be loosely referred as ``noisy'', they are probably more accurately described as ``noise-contaminated'' flow fields. Note that, in principle, measured magnetic fluctuations within a quiet-sun pixel can be due to physical evolution and not noise, but can nonetheless introduce spurious flow components. For example, sub-resolution fields in quiet-sun areas, which have significant field strength (on the order of hG, e.g., \citealt{rubio19}) but small filling factor, can produce measurable polarization within a sensitive enough instrument's pixel, and changes in this polarization can occur due to their evolution within a pixel.   This evolution is, however, inconsistent with the assumption inherent to optical flow methods, i.e., that changes in flux density arise due to flux transport from neighboring pixels.  Thus, although such sub-resolution magnetic evolution is physical and not due to measurement errors (such as CCD noise), its effect on flow estimation methods in quiet sun areas can be the same: an optical flow method will introduce spurious flow components to match the measured change.  Although of different origin, we will also refer to the effect of unresolved, rapidly fluctuating fields as ``noise''. 
In addition, LCT assumes that there is no horizontal magnetic field or that there is no vertical velocity (see, e.g., \citet{demoulin03}), i.e., LCT assumes an oversimplified equation for the evolution of the vertical flux.
Thus, methods such as the LCT cannot reconstruct flows with complete accuracy. The inferred flows are estimates and are, likely, noise-contaminated. 

Due to the aforementioned problems, a ``gap-filling'' or inpainting (the technique of modifying an image in an undetectable form \citep{bertalmio00}) method is required before the derived velocity fields can be used to infer the motion of passive scalars in the photosphere or be incorporated into coronal MHD simulations, e.g., to derive an electric field consistent with observations at the photospheric boundary. In the particular case of applying local correlation tracking (or other flow estimation methods) to magnetograms, estimated flows will be significantly noise-contaminated in pixels where the measured change in flux density (between initial and final frames) is not much larger than the measurement uncertainty in flux density.   In regions where the change in flux density is comparable to or smaller than measurement uncertainty, estimated flows will be worse -- not just contaminated, but ``noise-dominated''.  These conditions are typically met in regions outside active-region fields, where significant fields are present, but they are not spatially coherent. Consequently, we have focused on inpainting flows in such regions. Reconstruction of flows where spatially coherent magnetic fields are not present has potential applications for data-driven models of the solar atmosphere (e.g., \citealt{hoeksema20}).  The subject of appropriate choices of boundary conditions for dynamical models in weak-field regions is an area of ongoing research \citep{mackay21}.  Consequently, methods to inpaint flows in regions lacking strong, coherent magnetic fields are of interest. 

Inpainting of flows could be valuable in other contexts, too.  In fact, inpainting could be useful in any situation where information about flows over an entire region is sought, but flow tracers in remote-sensing observations (or sensing instruments in laboratory experiments) are sparse or non-existent in some sub-regions.  For instance, cloud motions have been used to infer velocities for weather forecasting (e.g., \citealt{horvath01}), but some areas are cloud-free, and inpainting could be useful in such areas.  Correlation tracking has also been applied to SOT prominence observation in the corona by \citet{freed16}, and to AIA post-flare arcades by \citet{freed18}, and in both cases there were areas with weak image intensities and therefore missing flow fields.  
For a recent review of inpainting techniques, see \citet{elharrouss20}.

In this letter, we show how a simple inpainting technique for highly corrupted images can be used to fill the gaps in noise-dominated velocity vector fields. Section \ref{sec method} describes the proposed Modified Monte-Carlo (MMC) method for image inpainting; section \ref{sec simulated} applies the MMC method to three-dimensional numerical simulations of the solar atmosphere; section \ref{sec observations} applies the MMC method to a velocity field derived from observational data of solar active region AR 10930; a discussion on the limitations of the methodology and conclusions are given in section \ref{sec conclusion}.

\section{Gap-filling method}
\label{sec method}

Image inpainting is a technique for restoration of an image with missing or corrupted points or regions. For relatively small damage, many inpainting algorithms, based on different approaches to the reconstruction, provide reasonable results (for a survey, see, e.g. \citealt{zarif15,tauber07,jam21}). It is more difficult to restore images containing large corrupted domains.  
There is no universal method that would provide good results for images and forms of corrupted areas of different types; in each particular case, an appropriate method needs to be found and its values of parameters should be carefully chosen.

In what follows, we deal with solar image data (shown below) containing both small and large corrupted regions, including extra-large corrupted areas with few non-corrupted pixels that are located far from each other. Therefore, we are forced to combine different approaches. Via many numerical experiments, we found that optimal results are obtained by a combination of two recovery methods, both based on a stochastic principle.
The horizontal velocity fields with missing data are treated as two images, one for each component, where the corrupted pixels coincide with the missing data. These images are reconstructed by a variant of the Monte-Carlo method described below, with non-missing velocity values kept intact. 

\subsection {The Standard Monte-Carlo (SMC) method \label{sec:SMC}}

From each corrupted point (pixel) of the image we start $n$ random walks simulating trajectories of a Wiener (white noise) process. Each trajectory is represented by a piecewise linear function constructed by the standard method: direction of the trajectory and its length are chosen randomly at each step of the random walk. Direction is parameterized by a random variable uniformly distributed on $[0, 2\pi)$ (polar angle); length is defined by the normal Gaussian distribution $\mathcal{N}(0, \sigma)$), value of the variance, $\sigma^2$, is a parameter of the method, in computations we used $\sigma = 0.5$ pixels. Once values for direction and length are randomly chosen, the trajectory advances in that direction by that distance and this procedure repeats. Each random walk continues until one of the two conditions is met: either the trajectory meets a non-corrupted point (such a trajectory is termed {\it successful}), or the number of steps exceeds a certain threshold $N$ (an {\it unsuccessful} trajectory). If the number of successful random walks started from a given corrupted point is large enough (at least $2n/3$ in our computations), then the corresponding corrupted point is assigned the intensity equal to the arithmetic mean of the intensities of all non-corrupted pixels met by the successful trajectories.
By the Feynman-Kac formula, the reconstructed intensities computed by the SMC method converge to a harmonic function: the solution of the Dirichlet problem for the Laplace equation in a certain domain \citep{gu04}. (Hence, this method can also be considered as a diffusion reconstruction method \citep{jam21}.) 

\subsection {The Modified Monte-Carlo (MMC) method \label{sec:MMC}}

If in the vicinity of a corrupted pixel there are no or few non-corrupted pixels, neither the Monte Carlo method described above, nor most other methods give good results of reconstruction. For example, when using the inpainting algorithm based on hypoelliptic diffusion \citep{boscain14}, large corrupted regions are not fully restored. Using the Averaging and Hypoelliptic Evolution (AHE) method for highly corrupted images \citep{boscain18, esaim18} the reconstruction is better, but still unacceptable. The main reason is the ``mosaic effect'' consisting in that large corrupted domains are reconstructed as regions of almost constant color (see Fig.~5 (step 1) in \cite{boscain18} and discussion therein). For large and very large corrupted regions this cannot be removed by the anisotropic diffusion at the next step of the AHE algorithm. In the present paper we use the method described in what follows, which we call the Modified Monte-Carlo method (MMC), which is a modification of the method presented in the previous subsection. 

The only modification is aimed at decreasing computational burden of the problem. It concerns the reconstruction of the pixels where the SMC does not provide enough information, i.e. the corresponding random walks are unsuccessful, therefore, more random walks and longer trajectories are required to be computed, making the SMC very demanding from the computational point of view. 

As for the SMC, in what follows we describe the algorithm for one corrupted pixel, $P_{km}=(x_k, y_m)$, assuming that the same procedure is repeated for all corrupted pixels independently. The pixel $P_{km}$ is surrounded by corrupted and non-corrupted pixels. Consider a  neighborhood of $P_{km}$: a square centered in $P_{km}$ defined as 
$$ U_r (x_k, y_m) = \{(x_i, y_j): \ | k-i | \le r, \ | m-j | \le r \},$$ 
containing $(1+2r)^2$ pixels, including $P_ {km}$ itself. 
If for the current pixel $P_{km}$ there exists a neighborhood $ U_r(x_k, y_m)$ of a relatively small size (we used $r\le 5$), containing more than a half non-corrupted pixels, we use the standard Monte-Carlo method (SMC) described in the previous section. Otherwise, many random walks are unsuccessful, making the standard Monte-Carlo method expensive from the  computational point of view, therefore, in order to reduce the execution time of our codes we use the procedure described below.

First, we consider the 9-point neighborhood $U_9 (x_k, y_m)$, containing $q\ge 5$ corrupted pixels (if $q<5$, the intensity of $P_ {km} $ is reconstructed by the SMC). Second, we increase the size of the neighborhood $U_9(x_k, y_m)$ until the number of non-corrupted pixels becomes at least $R=qM$, where $M$ is a parameter (we set $M = 5$ in computations). Let $G_R (x_k, y_m)$ be the set of non-corrupted pixels from the neighborhood $U_R(x_k, y_m)$.
Third, we randomly split (without replacement) the set of the non-corrupted pixels $G_R(x_k, y_m)$ into $q$ parts: $G^1, \ldots, G^q$, each  containing at least $M$ pixels, and for each part calculate the average intensity $F(G^i)$, $ i = 1, \ldots, q$; here $F(M)$ stands for average intensity (arithmetic mean) for all pixels in a set of non-corrupted pixels $M$. Finally, we randomly assign without replacement each corrupted pixel from the neighborhood $ U_9(x_k, y_m)$ to one of the values $F(G^1), \ldots, F (G^q)$.

As mentioned before, the non-corrupted pixels are not affected by both methods (SMC and MMC), in other words, in such points the given vector field remains intact.

\section{Analysis of simulated Data}
\label{sec simulated}

\begin{figure}[htp!]
\centerline{\includegraphics[width=1\columnwidth]{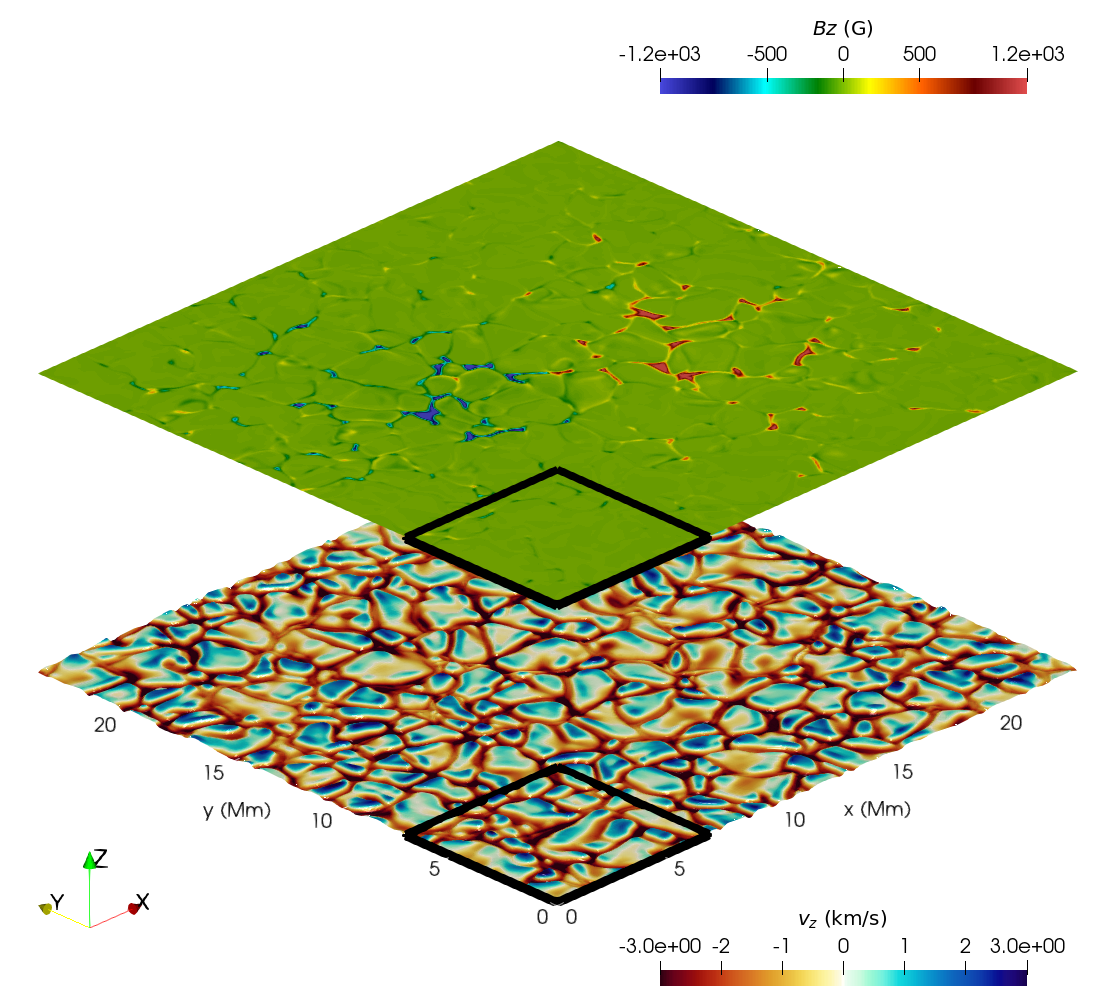}}
\caption{Bifrost simulated data of solar photosphere at $z=0$ and $t=3850$ s: vertical magnetic field, $B_z$ in Gauss (top) and vertical velocity field $u_z$ in km/s (bottom).}
\label{fig1}
\end{figure}

First, we illustrate our method with data obtained from a numerical simulation of the solar atmosphere.
We employ publicly available data from the 3D radiation magnetohydrodynamic code Bifrost, for simulating solar and stellar atmospheres. 
Bifrost uses a staggered grid and a 5th/6th order compact explicit finite difference scheme
with diffusive terms to ensure numerical stability. For detailed information on the code, see \cite{gudiksen11}.
We chose a simulation where the vertical domain extends from 2.4 Mm below the visible surface to 14.4 Mm above the surface, including the upper part of the convection zone, the photosphere, 
the chromosphere, the transition region, and the corona. 
The numerical grid has $504\times 504 \times 496$ points and represents a region of $24 \times 24 \times 17$ Mm$^3$ with 48 km for horizontal resolution, 
while the vertical resolution varies from 19 km in the photosphere and chromosphere to 100 km at the top boundary.
The data are in SI units, specifically, velocity is in m/s and magnetic field is in Tesla. 
The average unsigned magnetic field strength in the photosphere is 5 mT (50 G) with two dominant opposite polarity regions 8 Mm apart constituting an enhanced network. 
The full simulation data are available from the Hinode Science Data Centre Europe\footnote{http://www.sdc.uio.no/search/simulations}, under the name en24048\_hion. 
More details about this simulation are found in \cite{carlsson16}.
Figure \ref{fig1} shows the vertical components of the magnetic (top plane) and velocity (bottom plane) fields at $t=3850$ s and $z=0$, where the visible solar surface is defined.
It can be seen that the magnetic field is concentrated in the intergranular lanes and two large, opposite-polarity regions are present. The magnetic and velocity field units have been converted to Gauss and km/s, respectively.

\begin{figure}
\centerline{\includegraphics[width=1\columnwidth]{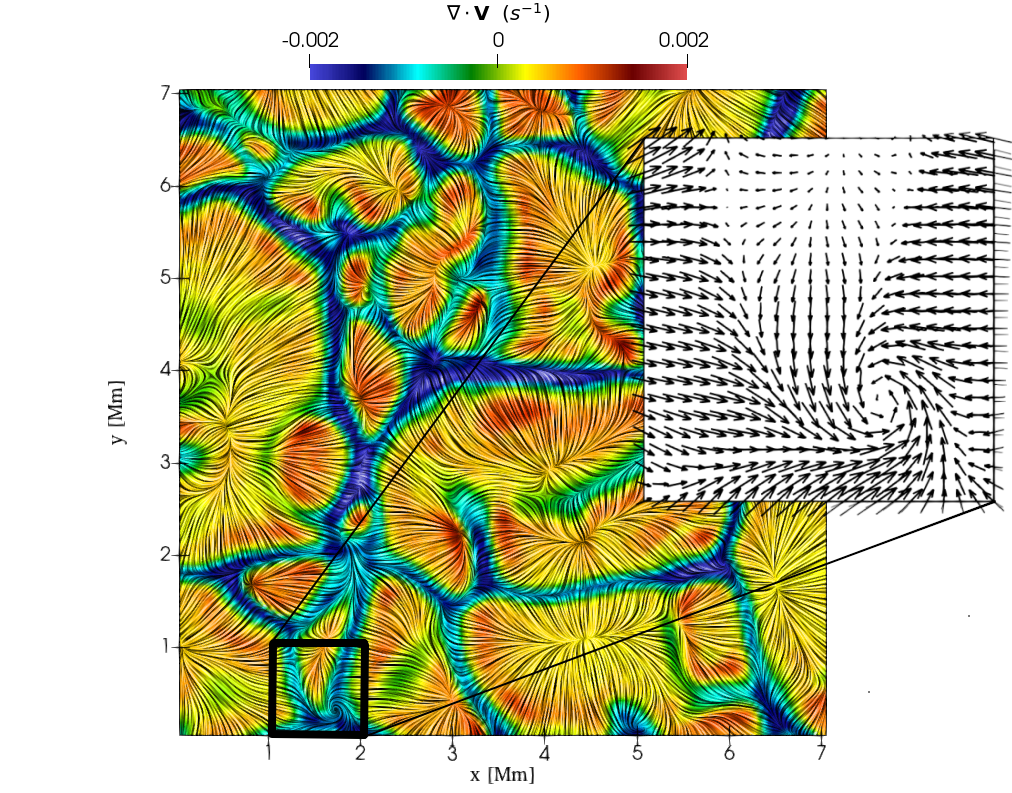}}
\centerline{\includegraphics[width=1\columnwidth]{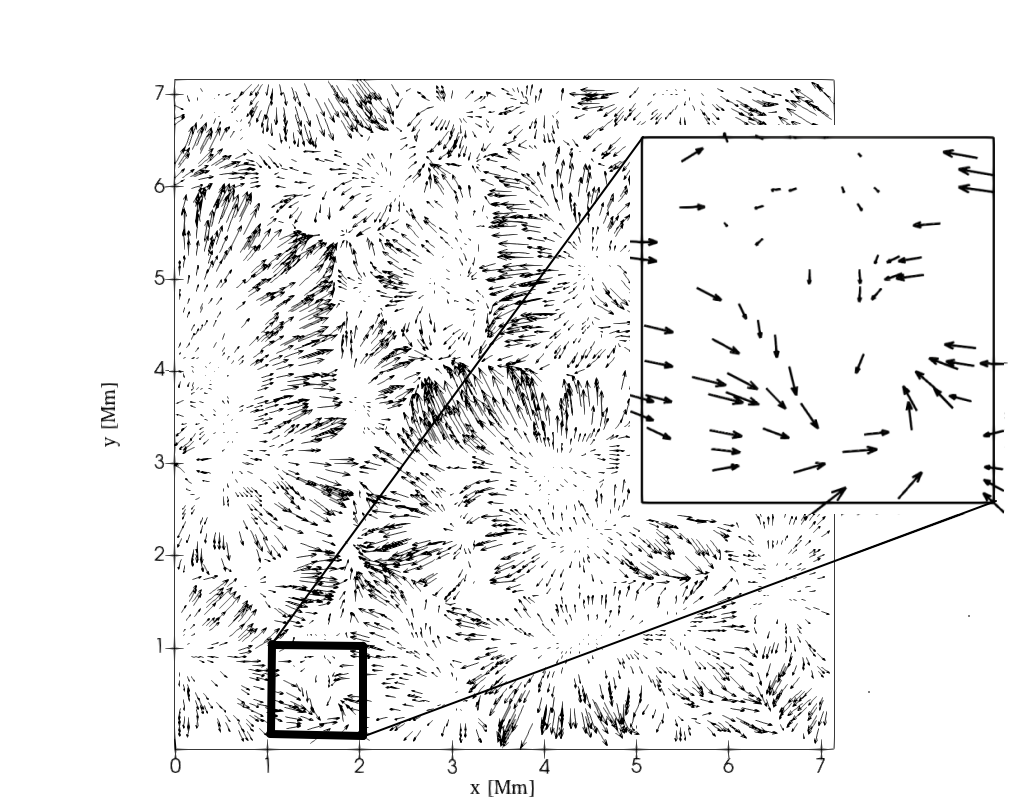}}
\centerline{\includegraphics[width=1\columnwidth]{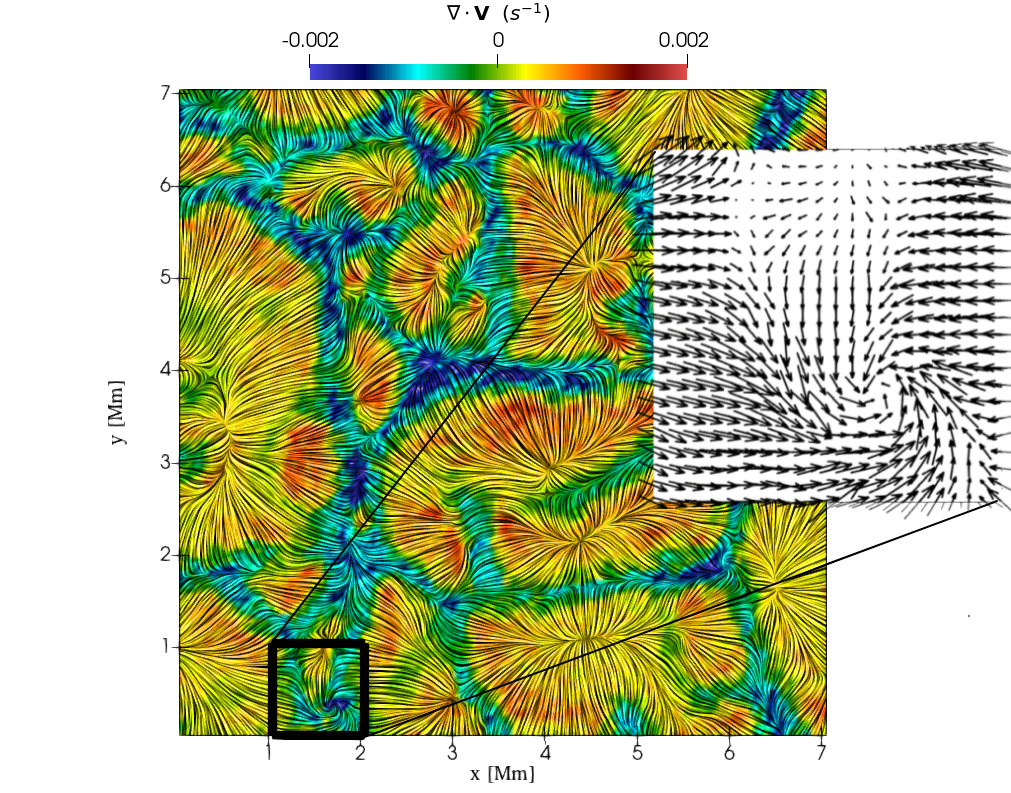}}
\caption{ Simulated solar atmosphere at 150 km above the solar surface: original Bifrost data (top panel), original velocity field data with random gaps (middle panel), and reconstructed velocity field, gap-filled by the MMC method (bottom panel). In the upper and lower panels, the line integral convolution (LIC) of horizontal velocity field vectors is colored by $\nabla \cdot \mathbf{v}$. In the middle panel  and in the enlarged views the velocity field is represented by arrows.}
\label{fig2}
\end{figure}

\begin{figure*}
\centerline{\includegraphics[width=2\columnwidth]{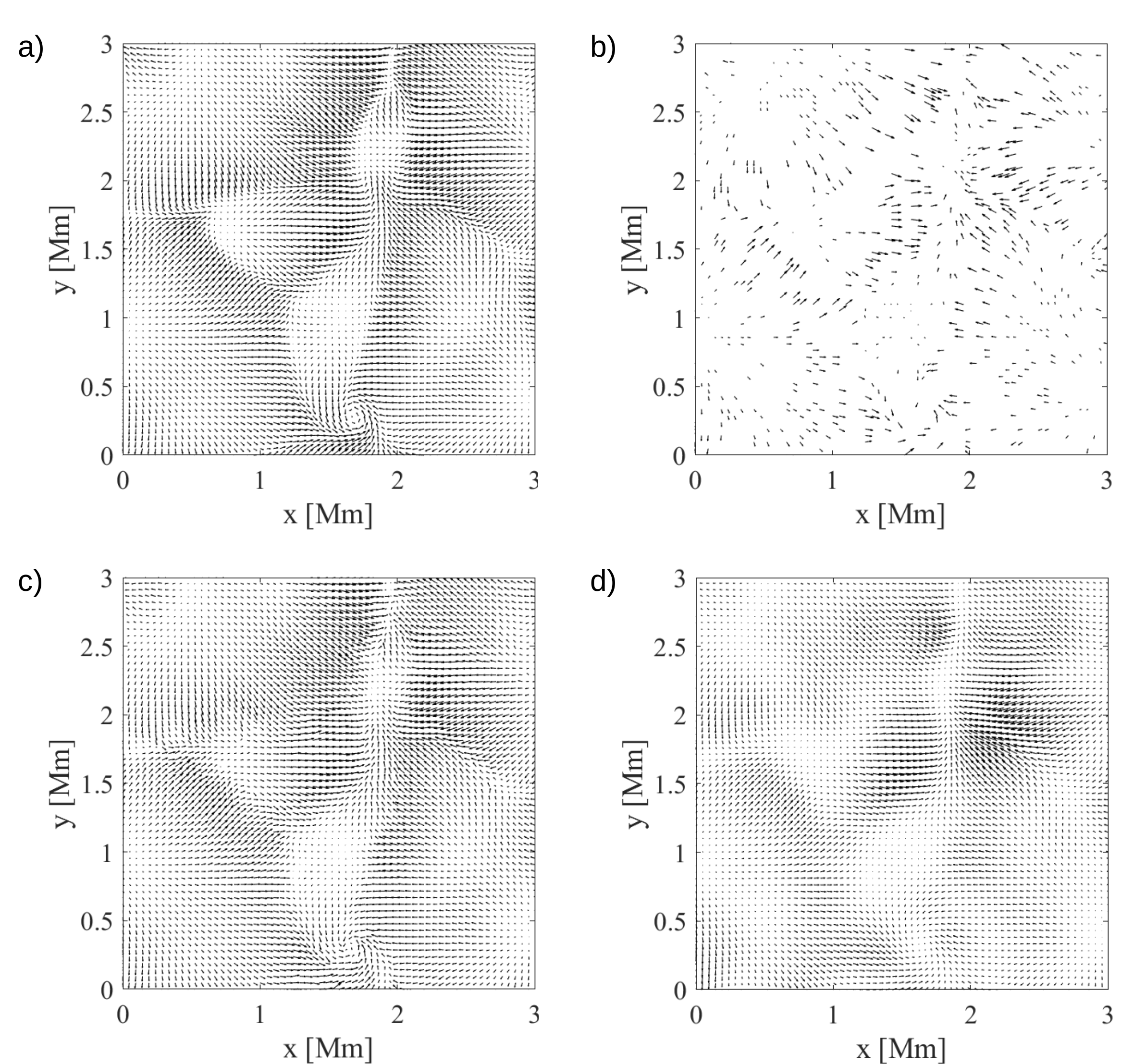}}
\caption{ Comparison between the MMC method and a smoothing routine: original Bifrost velocity field data (a); original velocity field data with random gaps (b); reconstructed velocity field, gap-filled by the MMC method (c); reconstructed velocity field, gap-filled by the smoothing method (d).}
\label{fig2b}
\end{figure*}

Since we plan to compare simulated data with satellite data, we are only interested in the horizontal components of the velocity field 
on the photosphere. Thus, we select a 2D slice in the box shown in the lower corner of the planes in Fig. \ref{fig1} to illustrate the method. 
The corresponding streamlines, colored by the divergence of the horizontal velocity field, are shown in the upper panel of Fig. \ref{fig2}. The divergence was computed using second-order, centered finite differences. The inset displays an enlargement of a box near the lower-left corner, with velocity vectors surrounding a vortex structure. In order to test our inpainting  method, we first produce a corrupted velocity field from this set by randomly removing vectors from it. For each vector position, a random variable is generated from a Gaussian distribution with zero mean and variance equal to 50; if the random variable so generated has absolute value larger than 10, the vector in that location is removed. The resulting vector field has $\approx 84\%$ of the original vectors removed and is shown in the middle panel of Fig. \ref{fig2}. Once again, the inset shows an enlargement of the small box at the bottom, where the frequency of gaps in the corrupted image can be appreciated. The bottom panel shows the streamlines and divergence of the restored (gap-filled) velocity field obtained by the MMC algorithm, which is visually very similar to the original one. A closer look at the small box, shown in the inset, attests the power of the method to rebuild a vector field from a set of a few scattered vectors. The Pearson correlation coefficient \citep{fisher58}
between the matrix of $x$-components of the original velocity field and the matrix of $x$-components of the gap-filled velocity field is 0.98, the same value obtained for the correlation involving the matrices of $y$-components.
The correlation coefficient between original and gap-filled divergence fields is 0.88. 
As a comparison, we performed the inpainting of the same data using the discrete cosine transform with penalized least squares (DCT-PLS) method, a popular smoothing technique introduced by \citet{garcia10}.
The method is capable of handling large areas of missing values and has been extensively used in the literature (see, e. g., \citealt{wang22}). 
The automatic choice of the amount of smoothing is performed by minimizing the generalized cross-validation score and a Matlab code is provided in \citet{garcia10}.
The results are summarized in Fig. \ref{fig2b}, where (a) shows the lower-left part of the domain with the original Bifrost velocity field, (b) shows the velocity field with gaps, as in the middle panel of Fig. \ref{fig2}, (c) shows the velocity field inpainted by the MMC method and (d) shows the velocity field inpainted by the DCT-PLS method. Note that the DCT-PLS procedure removes many of the small details and sharp gradients present in the original field, as expected for a smoothing method. The correlation coefficient between the DCT-PLS and the original field is 0.94, a little smaller than the one obtained with the MMC method (0.98). Our goal is not to conduct an extensive comparison with this smoothing procedure and we don't claim that our method is better for all applications. We want to stress that for the inpainting of a two-dimensional field extracted from a three-dimensional system, a smoothing procedure may lose some of the fine details and sharp gradients observed in the original field, which is something that the MMC recovered quite well\footnote{Note that the solution of the SMC method is ``smooth" in the sense of elliptic regularity. However, this is different from the ``smoothing" of the DCT-PLS method, which means  that small scales are filtered out because they are assumed to represent noise in the data.}. In a future work, we also intend to explore the robustness of the method as a function of the noise level, but for now, we conclude that the MMC gap-filling method proposed in this letter is accurate for this task and proceed to employ it with real observational data.

\section{Analysis of Satellite data}
\label{sec observations}

The photospheric horizontal velocity field is estimated from solar line-of-sight magnetograms using
the Fourier local correlation tracking (FLCT) method \citep{welsch04}. 
To obtain the magnetograms, we used Stokes V/I from Hinode/NFI (Narrowband Filter Imager)
observations in Fe I 6302 {\AA} of solar active region AR 10930 on 12 December 2006. 
The noise level was estimated at $\approx$17 G by fitting the core of histogrammed field strengths \citep{hagenaar99}. 
Considering the reduction of noise due to the averaging in the tracking procedure, a 
tracking threshold of 15 G was chosen, meaning that no velocities are assigned to magnetogram pixels below this threshold. 
The windowing parameter, $\sigma$, used by FLCT was set to 4 pixels.
The cadence of the magnetogram images is $\approx$121 s and
the sampling time between velocity field frames is $\Delta t = 8$ min.  
This is small enough to minimize decorrelation between frames, while allowing for boxcar averaging of 5 
magnetograms to produce each velocity frame, which reduces noise significantly.
Calculations with $\Delta t = 4$ min result in qualitatively similar results.
For a thorough description of how the FLCT method was fine--tuned for this problem, see 
\cite{welsch12}. For other works on the same velocity field, see \cite{yeates12} and \cite{chian14}.

\begin{figure}
\centerline{\includegraphics[width=1\columnwidth]{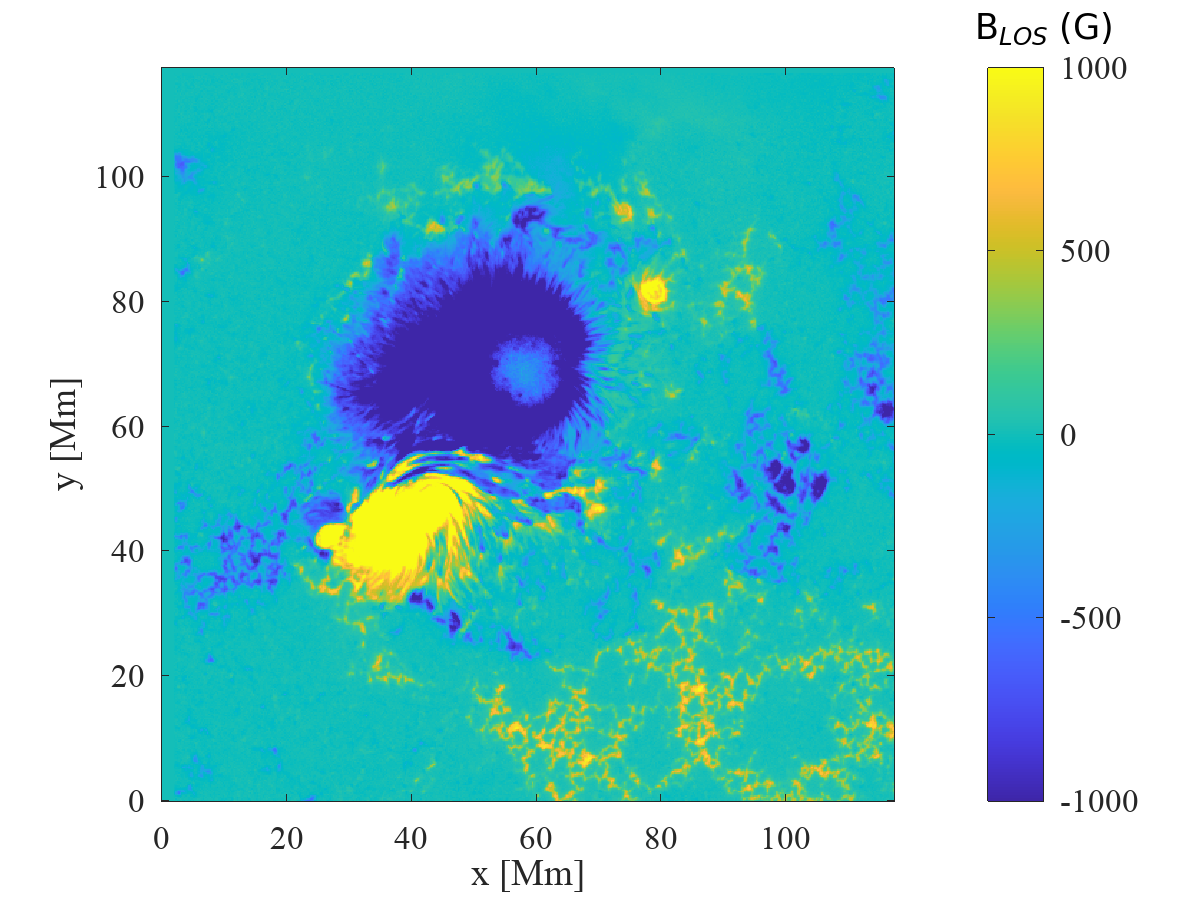}}
\centerline{\includegraphics[width=1\columnwidth]{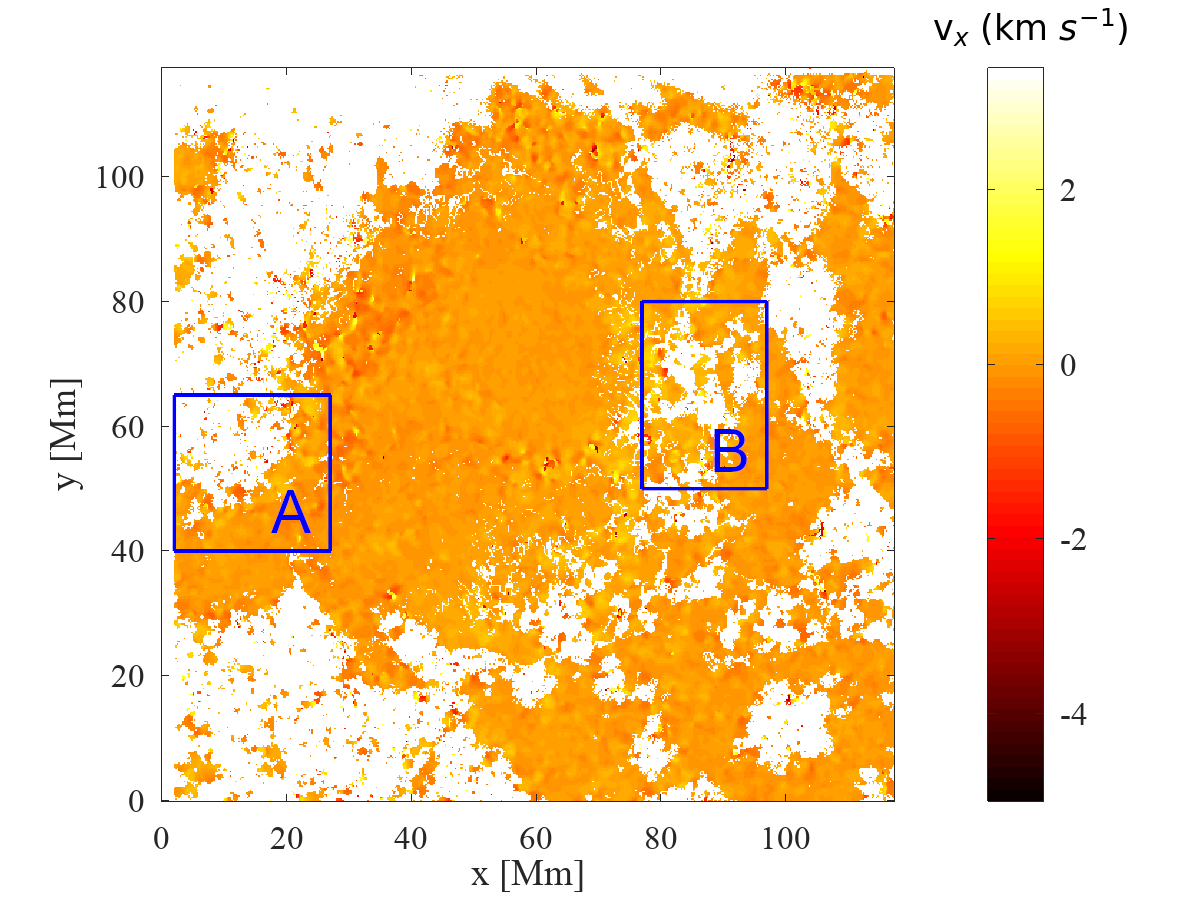}}
\caption{Original Hinode observations of AR 10930. Top: $B_{LOS}$ in Gauss; bottom: $v_x$ component of the velocity field in km/s, obtained from FLCT. White areas in the $v_x$ map represent gaps in the FLCT field due to below-threshold weak magnetic field values. }
\label{fig3}
\end{figure}

Figure \ref{fig3} shows the Hinode line-of-sight magnetogram of AR 10930 (top panel) for 17:20:44.525 UT on 12 December 2006 and the corresponding $x$ component of the velocity field obtained by the FLCT method (bottom panel). The line-of-sight magnetic field is in Gauss and the velocity field is in km/s.  (The apparent weak field in the negative sunspot's core is an artifact of our weak-field, linear calibration, which is inaccurate in strong-field regions. Absolute calibration of field strengths in sunspots is irrelevant for our purposes because the regions investigated in this study do not include umbral fields).  
The white values in the $v_x$ map represent gaps in the FLCT field. It can be seen that a considerable portion of the domain is void of velocity vectors. The two boxes marked as A and B indicate the regions where the velocity field gaps will be filled by the MMC method.
\citet{welsch15} compared the NFI ``line-wing'' magnetic flux densities and SP (full Stokes inversion, fill-fraction corrected) field strengths, and found an approximately linear scaling between the two for pixel-averaged flux densities up to about 1.5 kG, appropriate for the non-umbral fields in areas A \& B.  Consequently, although the NFI flux densities were not calibrated, NFI image gradients are expected, statistically, to be proportional to magnetic variations.

Regarding the use of LCT methods to reconstruct photospheric velocity fields from magnetograms, as noted by \citet{demoulin03}, either horizontal or radial flows (or a combination of these) can cause magnetic footpoints to move, so the apparent footpoint velocities do not correspond one-to-one to plasma velocities. In particular, this velocity will
not be captured by LCT specifically, as LCT does not utilize all three components of
the magnetic field to estimate the plasma velocity. Consequently, LCT flows estimated from magnetograms will not necessarily correspond to plasma flows. 
If the solar observations are not near the disk
center, the motion of plasma recorded in images is actually a projection of three-dimensional motions. 
We remark that AR 10930 in Fig. \ref{fig3} was very near disk center. In its Solar Region Summaries for 12 \& 13 December 2006, NOAA lists AR 10930 at coordinates S05W07 \& S06W21, so at the time of the data analyzed in the paper, circa 17:20 UT, the region's center was likely near S06W18.  Since projection effects scale as the cosine of viewing angle, we expect this should introduce errors of at most 10\%. 
One of the tracked regions analyzed, area A in Fig. \ref{fig3}, contains mostly plage fields, which are predominantly vertical. Along with the Hinode/NFI data, one Hinode/SP vector field map is available for 12 Dec, albeit after 20:00UT on 12 Dec. 2006 -- about three hours after the image shown in Fig. \ref{fig3}.  Area A is nearly the same area analyzed by \citet{welsch15}, who reported mean and median inclinations of less than 30 degrees from the radial direction (inward) from the SP data.  For this region, the effect of radial flows is expected to be small compared to transverse flows, since the radial direction is nearly along the field.   Consequently, in area A, we expect good correspondence between LCT-inferred motions and horizontal plasma velocities.  
On the other hand, area B's left edge lies at the western (rightward) edge of some penumbral fields, which are mostly horizontal, and also contains mixed polarity regions.  Analysis of the corresponding area in the SP map three hours later shows that this area's fields also tend to be vertical, with a median tilt of 23 degrees in pixels with $|B_z| > 15$.   Some substantially inclined fields are present, though, so LCT flows will not correspond to plasma velocities in some areas.  

\begin{figure*}
\begin{center}
{\includegraphics[width=0.45\textwidth]{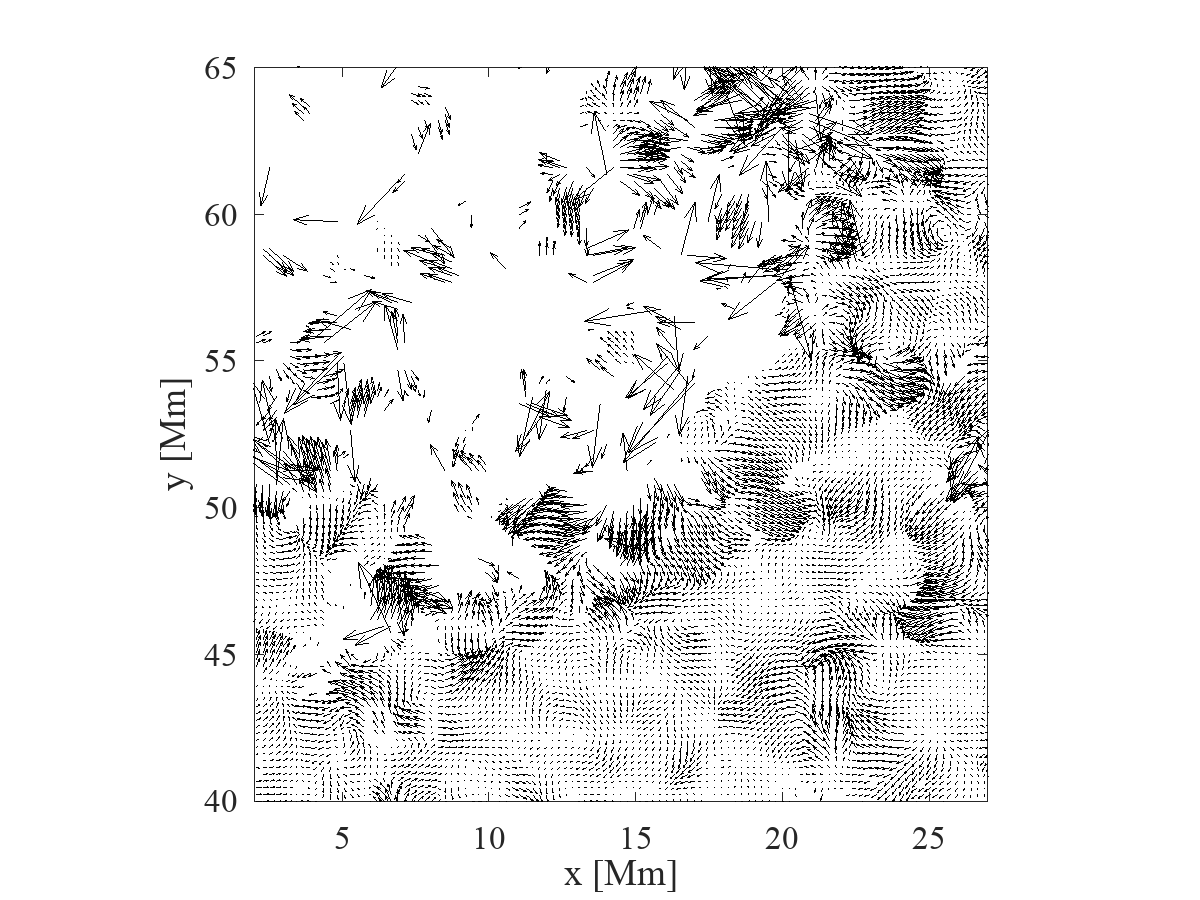}}
{\includegraphics[width=0.25\textwidth]{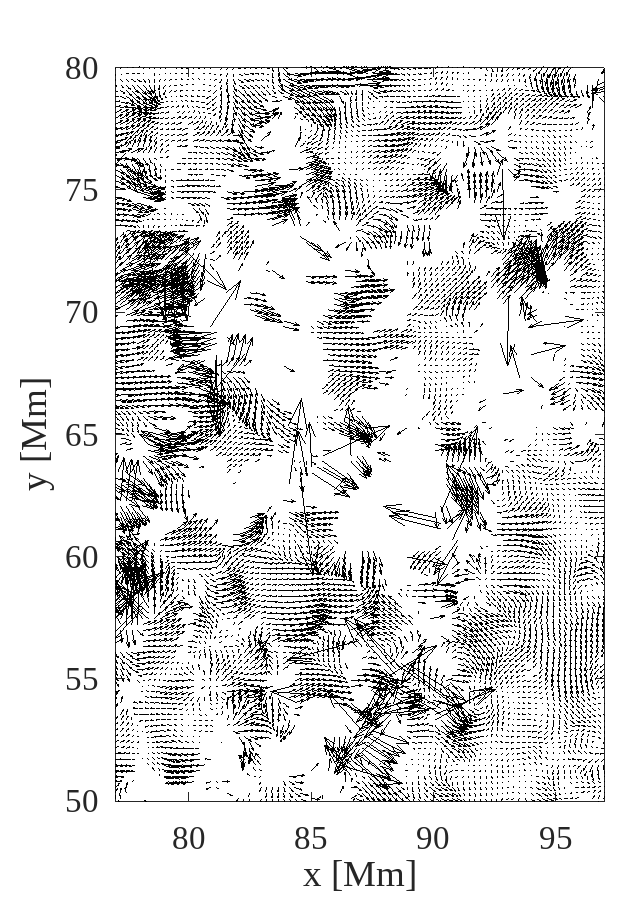}}\\
{\includegraphics[width=0.45\textwidth]{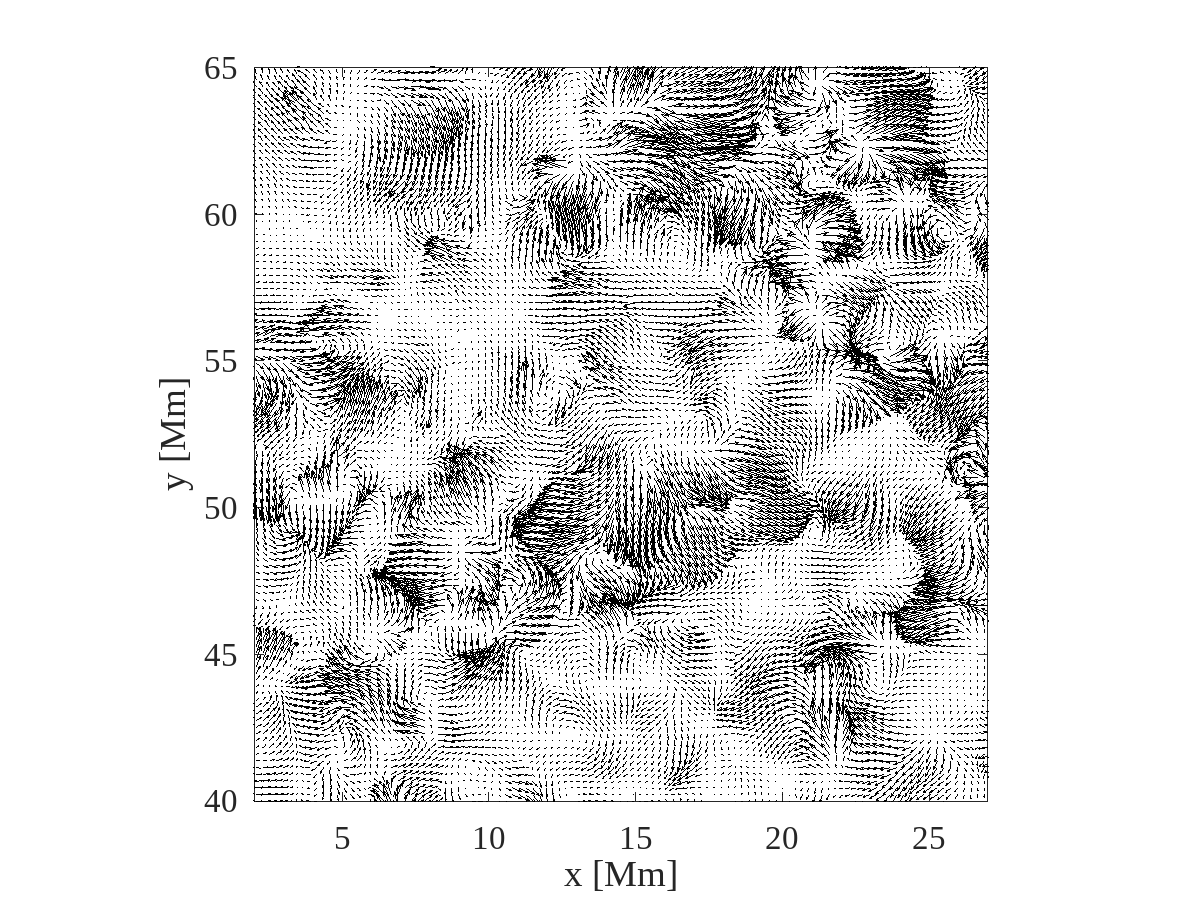}}
{\includegraphics[width=0.25\textwidth]{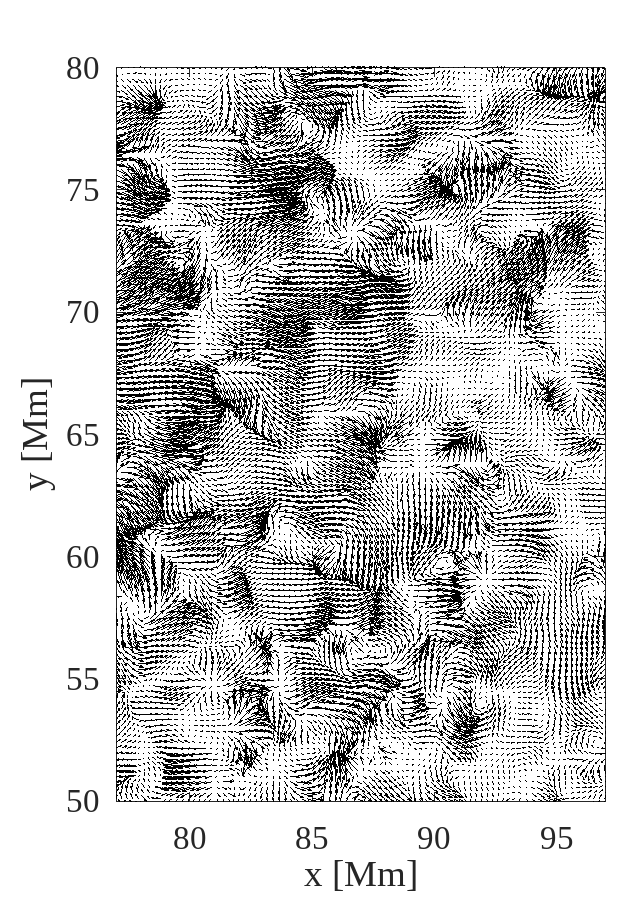}}\\
{\includegraphics[width=0.45\textwidth]{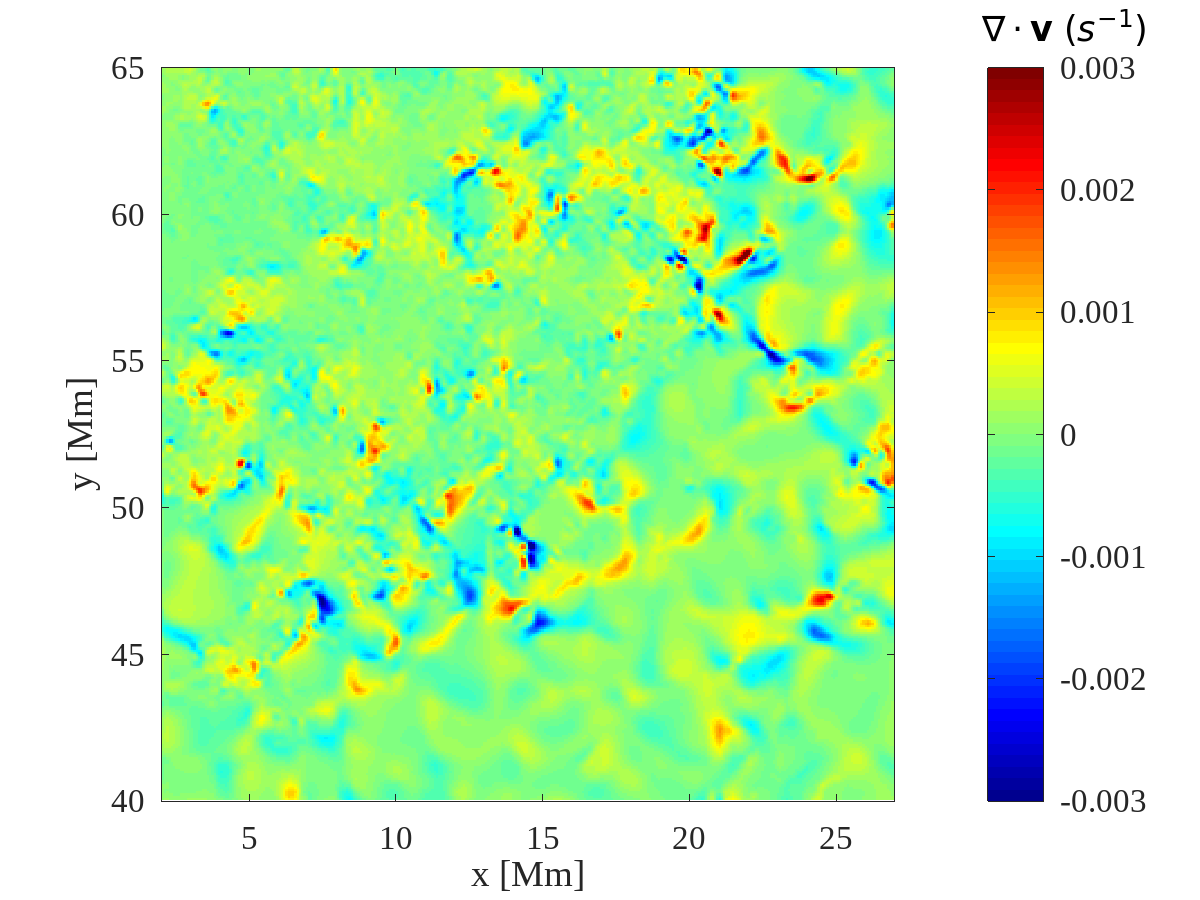}}
{\includegraphics[width=0.33\textwidth]{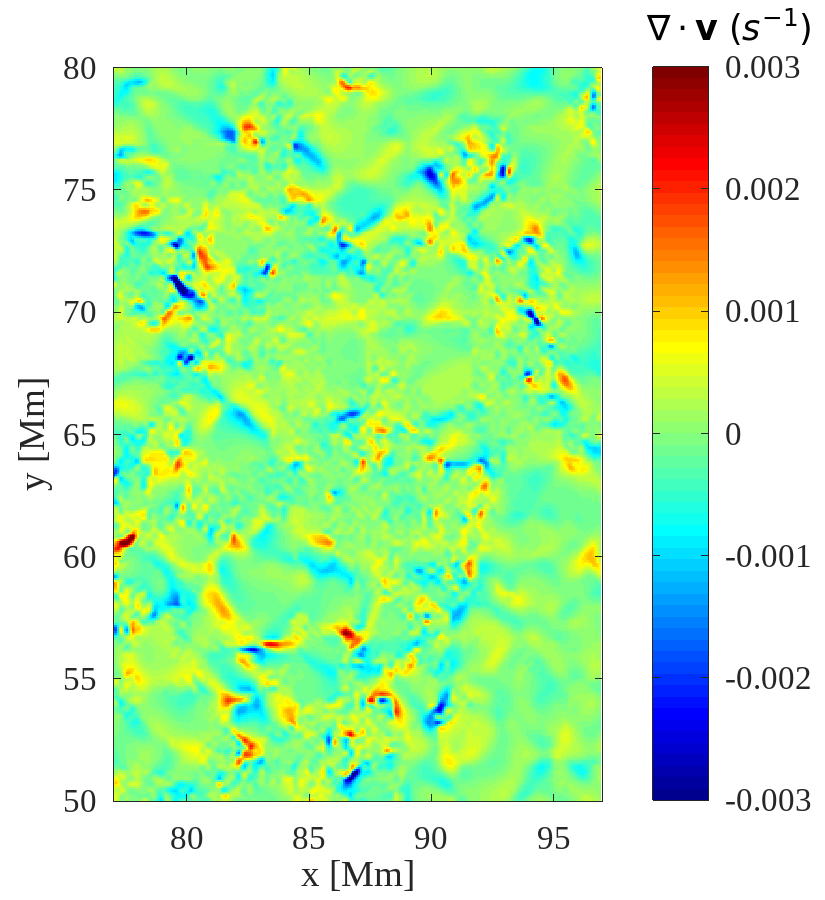}}
\end{center}
\caption{Reconstruction of Hinode velocity fields in AR 10930. Top: horizontal velocity field obtained from the FLCT method in boxes A (left) and B (right) of Figure \ref{fig3}; middle: MMC gap-filled velocity field; bottom: the divergence of v from the reconstructed MMC field.}
\label{fig4}
\end{figure*}

Figure \ref{fig4} illustrates the application of the MMC method in the subregions A and B indicated in the bottom panel of Fig. \ref{fig3}. The left panels correspond to region A and the right panels to region B.
The top panels of Fig. \ref{fig4} show the horizontal components of the original FLCT velocity field. Note that some FLCT vectors are much larger than average, and that essentially all occur at the edges of tracked regions.  These apparently large flows are unphysical, and 
occur in weak-field regions, where there is little genuine magnetic structure but much noise. 
These inaccurate flows are precisely why a field-strength threshold is used to determine which pixels should be tracked --- i.e., they are the reason gaps in the FLCT flow maps exist.
Thus, before applying the MMC method, we first eliminate all vectors for
which the modulus of one of the components is larger than 0.28 km/s. This threshold corresponds to four times the 
variance 
of the distribution of velocity field components. 
After the cleansing of spurious vectors, the MMC method is applied and the resulting field is seen in the middle panels of Fig. \ref{fig4}.
It is remarkable that even for such a sparse matrix of velocity vectors the method is still able to generate a field with fine structures that
seem to be coherent with what should be expected for that region. The accuracy of the reconstruction cannot be assessed for the observational data, since the real field is not available for comparison in those gaps. Because we have not yet undertaken the needed tandem tests of LCT flow estimation from simulated magnetic field data plus inpainting, we do not perform any quantitative analysis of the inpainted velocities here, and  make only qualitative comments.  We defer quantitative analysis of inpainted flows to future studies, which will involve further tests. 
Thus, the validity of inpainted flows inferred from magnetogram-derived flows has not been demonstrated in the current work. Our results obtained with the numerical simulations, however, encourage further exploration of the method in future studies. Because of systematic errors in the LCT estimates, it is reasonable to expect that the inpainted flows here are less accurate than those in our validation study using simulated data.
The bottom panels in Fig. \ref{fig4} depict the divergence field computed from the restored velocity field. 

\section{Conclusions}
\label{sec conclusion}

We have demonstrated that the Modified Monte-Carlo method introduced in this letter is a powerful
tool for reconstructing highly corrupted photospheric velocity fields. It can fill wide, contiguous areas of missing data while keeping the original vectors intact. Our method is much simpler 
than alternative image completion techniques based on artificial intelligence/deep learning. Such methods have been successfully applied to the restoration of images of global positioning system (GPS) measurements of the ionosphere \citep{chen19,pan20}, as well as solar images corrupted by flares \citep{yu21}. Deep learning techniques usually rely on training of a set of artificial neural networks using reference data before the networks can be used to fill the gaps in real observations. The training images must be provided by other observations or by numerical simulations, a step that is unnecessary in our method. 
Applications of the MMC method are in no way restricted to solar physics, as it may be readily applied to image restoration in general.

Regarding our analysis of simulated data, we note that our validation study was performed directly on the simulations' velocity fields, instead of on velocity fields estimated by LCT (or similar methods). Because velocities estimated by optical flow methods will, in general, contain inaccuracies, we expect that the accuracy of inpainted velocities will be degraded from the values we report here.  Accordingly, therefore, the tests performed here give ``best-case'' results.  Consequently, a ``tandem'' test of the combination of flow estimation plus inpainting results would be necessary to assess the overall accuracy.  Since the primary focus of this paper is the inpainting method, and not any particular flow estimation method with which it might be paired, we defer any tandem flow estimation plus inpainting tests to future studies.
The inpainting approach is separate from use of LCT, and could be coupled with other methods of determining velocity fields, such as balltracking \citep{potts03}.
Users who wish to apply the inpainting method in different contexts (e.g., different spatial resolution, or different data types) should be aware that the accuracy of reconstructions can differ substantially from the accuracy of our reconstructions for the particular simulated data analyzed here.
 
In this paper, we have: (1) introduced the idea of inpainting photospheric velocity fields, (2) described one approach, (3) tested this method with simulations, and (4) shown example results obtained from application to observations. While much further testing of this technique is warranted, it shows promise as a useful method to fill gaps in reconstructed photospheric velocity fields.

\section*{Acknowledgments}
ELR acknowledges Brazilian agencies CAPES (Grant 88887.309065/2018-01) and CNPq (Grant 306920/2020-4) for their financial support, as well as FCT -- Funda\c c\~ao para a Ci\^encia e a Tecnologia (Portugal). RC gratefully acknowledges AL ARISE (Ref. LA/P/0112/2020), SYSTEC Base (Ref. UIDB/00147/2020) and Programmatic (Ref. UIDP/00147/2020) funding by the Foundation for Science and Technology (FCT), Portugal. A part of computations was carried out in the framework of the computational project 2021.09815.CPCA financed by FCT and executed on the OBLIVION Supercomputer (based at the High Performance Computing Center, University of Evora) funded by the ENGAGE SKA Research Infrastructure (Ref. POCI-01-0145-FEDER-022217 – COMPETE 2020 and FCT) and by the BigDataUE project (Ref. ALT20-03-0246-FEDER-000033 – FEDER and the Alentejo 2020 Regional Operational Program). BTW acknowledges NASA LWS Award 80NSSC19K0072, and a grant from the Japan Society for the Promotion of Science that enabled estimating flow fields in AR 10930.

\bibliography{refs}{}
\bibliographystyle{aasjournal}

\end{document}